\documentclass[prb,twocolumn,superscriptaddress,showpacs,amsmath,amssymb]{revtex4}
\usepackage{float}
\usepackage{amsmath}
\usepackage{graphicx}
\usepackage{hyperref}
\usepackage{bm}
\usepackage[usenames]{color}
\usepackage{verbatim}
\newcommand{\be}{\begin{equation}}
\newcommand{\ee}{\end{equation}}   
\newcommand{\bea}{\begin{eqnarray}}
\newcommand{\eea}{\end{eqnarray}}
\newcommand{\phrl}[1]{Phys.~Rev.~Lett. {\bf #1}}
\newcommand{\phrb}[1]{Phys.~Rev.~B {\bf #1}}

\newcommand{\jpsj}[1]{J.~Phys.~Soc.~Jpn. {\bf #1}}
\newcommand{\RMP}[1]{Rev.~ Mod.~ Phys. {\bf #1}}
\newcommand{\jpcm}[1]{J.~Phys.:Condens.~Matter.{\bf #1}}
\newcommand{\bib}{\bibitem}
\newcommand{\lb}{\left[}
\newcommand{\rb}{\right]}
\newcommand{\lp}{\left(}
\newcommand{\rp}{\right)}
\newcommand{\lf}{\left\{}
\newcommand{\rf}{\right\}}

\newcommand{\p}{\mathbf{p}}

\renewcommand{\k}{\mathbf{k}}
\newcommand{\bk}{\widetilde{k}}
\newcommand{\bOmega}{\widetilde{\Omega}}
\newcommand{\bmu}{\widetilde{\mu}}
\newcommand{\bb}{\widetilde{b}}
\newcommand{\bepsilon}{\widetilde{\epsilon}}
\newcommand{\brho}{\widetilde{\rho}}

\newcommand{\Y}{\mathfrak{Y}}

\begin{document}

\title{Transport and optics at the node in a nodal loop semimetal}

\author{S. P. Mukherjee}
\affiliation{Department of Physics and Astronomy, McMaster University, Hamiltion, Ontario, Canada L8S 4M1}

\author{J. P. Carbotte}
\affiliation{Department of Physics and Astronomy, McMaster University, Hamiltion, Ontario, Canada L8S 4M1}
\affiliation{Canadian Institute for Advanced Research, Toronto, Ontario, Canada M5G 1Z8}

\begin{abstract}
We use a Kubo formalism to calculate both A.C. conductivity and D.C. transport properties of a dirty nodal loop semimetal. The optical conductivity as a function of 
photon energy $\Omega $, exhibits an extended flat background $\sigma^{BG}$ as in graphene provided the scattering rate $\Gamma$ is small as compared to the radius of 
the nodal ring $b$ (in energy units). Modifications to the constant background arise for $\Omega\le \Gamma $ and the minimum D.C. conductivity $\sigma^{DC} $ which is 
approached as $\Omega^2/\Gamma^2$ as $\Omega\rightarrow0$, is found to be proportional to $\frac{\sqrt{\Gamma^2+b^2}}{v_{F}}$ with $v_{F}$ the Fermi velocity. For $b=0$ 
we recover the known three-dimensional point node Dirac result $\sigma^{DC}\sim \frac{\Gamma}{v_{F}}$ while for $b>\Gamma$, $\sigma^{DC}$ becomes independent of $\Gamma$ 
(universal) and the ratio $\frac{\sigma^{DC}}{\sigma^{BG}}=\frac{8}{\pi^2}$ where all reference to material parameters has dropped out. As $b$ is reduced and becomes of 
the order $\Gamma$, the flat background is lost as the optical response evolves towards that of a three-dimensional point node Dirac semimetal which is linear in $\Omega$ 
for the clean limit. For finite $\Gamma$ there are modifications from linearity in the photon region $\Omega\le \Gamma$. When the chemical potential $\mu$ (temperature $T$) is nonzero the D.C. conductivity increases as $\mu^2/\Gamma^2$($T^2/\Gamma^2$) for $\frac{\mu}{\Gamma}$ $\lp\frac{T}{\Gamma}\rp \le 1$. 
Such laws apply as well for thermal conductivity and thermopower with coefficients of the quadratic law only slightly modified from their value in the three-dimensional 
point node Dirac case. However in the $\mu=T=0$ limit both have the same proportionality factor of $\sqrt{\Gamma^2+b^2}$ as does $\sigma^{DC}$. Consequently the Lorentz 
number is largely unmodified. For larger values of $\mu>\Gamma$ away from the nodal region the conductivity shows a Drude like contribution about $\Omega\approxeq 0$ 
which is followed by a dip in the Pauli blocked region $\Omega \le 2\mu$ after which it increases to merge with the flat background (two-dimensional graphene like) for 
$\mu< b$ and to the quasilinear (three-dimensional point node Dirac) law for $\mu> b$.

\end{abstract}


\maketitle

\section{Introduction}
\label{sec:I}

Optical (IR) along with other spectroscopies such as angular resolved photo emission ARPES and scanning tunelling microscopy have given us a wealth of information on the
dynamics of charge carriers in metals and superconductors \cite{Carbotte, Basov, Jiang} with different gap symmetries. More recently the dynamic optical conductivity 
$\sigma(T, \Omega)$ as a function of temperature and photon energy has been equally successful when applied to the class of two-dimensional (2D) metals such as graphene 
\cite{Li,Stille, Sharapov}, the surface states of topological insulators \cite{Schafgans,Zhou} as well as topological materials such as Dirac and Weyl semimetals 
\cite{Chen,Sushkov,Xu,Neubauer,Timusk,Chinotti,Tabert,Nicol,Carbotte1}.

Another recent development has been the discovery of nodal loop semimetals \cite{Burkov,Aji,Carter,Kim,Weng,Uchoa,Fang,Bian,Yamakage,Araújo,Ezawa}. Their magnetic 
susceptibility \cite{Koshino}, density fluctuation plasmons and Friedel oscillations\cite{Rhim}, Landau quantization \cite{YBKim} and some aspect of their topological 
electrodynamic response \cite{Ramamurthy} have been studied. The dynamical optical conductivity as a function of photon energy $\Omega$ has also been considered in the 
clean limit \cite{Carbotte2}. It was found to display signatures of both three-dimensional (3D) point node Weyl or Dirac-like materials and 2D graphene-like systems, 
depending on what range of photon energy $\Omega$ is used to probe the dynamics. For $\Omega$ small compared with twice the radius in energy units of the nodal ring($b$),
the response is 2D in nature while for $\Omega>2b$ it evolves to 3D, characteristic of point node Dirac. In any realistic case, the charge carriers will also have a 
finite scattering rate $\Gamma$ which influences their motion. In this paper we study the effect of $\Gamma$ on the electromagnetic properties of a nodal loop semimetal 
at finite chemical potential and photon energy. In addition, we consider D.C. transport including conductivity, thermal conductivity, thermopower and Lorentz number. Here 
we will be particularly but not exclusively interested in the case when the chemical potential $\mu$ and temperature T are small compared with $\Gamma$ which allows 
optics and transport at the nodes to be probed. This regime includes the concept of minimum conductivity and how it is modified when $\mu$ and/or T is increased out of 
zero.

In section II we present the necessary formalism including the Kubo formula for the dynamical conductivity $\sigma(T,\Omega)$ at finite T and photon energy. Results at 
finite photon energy are given in section III. While many of our results are for $\mu=0$ (nodal region) the effect of a finite chemical potential are also presented. In
section IV we consider D.C. properties, electrical conductivity, thermal conductivity, thermopower and Lorentz number. A discussion and conclusion can be found in 
section V.

\section{Formalism}
\label{sec:II}

The continuum $4\times 4$ matrix Hamiltonian for a loop node semimetal on which our work is based takes the form
\be
\label{Basic Hamiltonian}
\hat{H}=v_{F} \hat{\tau}_x \lp \hat{\mathbf{\sigma}}.\p\rp + b \hat{\tau}_z \hat{\sigma}_x,
\ee
where $v_{F}$ is the Fermi velocity, $\p$ is the momentum equal to $\k$ and b is a Zeeman field oriented along the $x$-axis. The $2\times 2$ matrix $\hat{\tau}$
and $\hat{\mathbf{\sigma}}$ are each a set of Pauli matrices. For convenience in our calculation we will set $v_{F}=\hbar=1$ and only at the end restore them. The 
energies of the two sets of bands involved can be written as,
\begin{figure}[h]
\vspace{-1.5cm}
\hspace*{-4.0cm}
\includegraphics[width=3.5in,height=3.6in, angle=0]{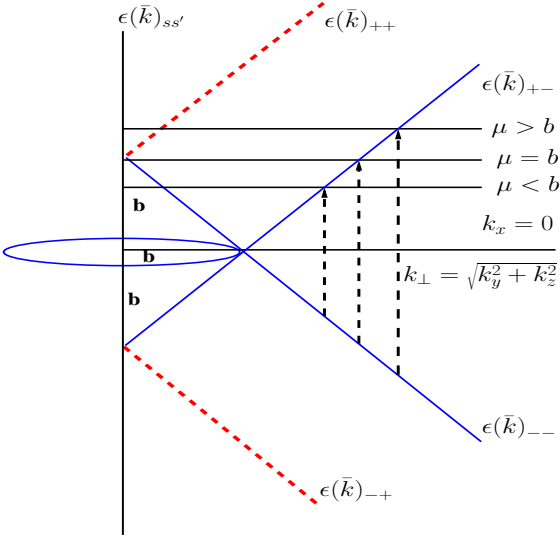} 
\vspace{-0.5cm}
\caption{(Color online) Schematic diagram showing the electron dispersion curves $\epsilon_{ss'}(\k)=s\sqrt{k^2_x+\lp\sqrt{k^2_y + k^2_z}+s'b\rp^2}$. In the figure
$k_{x}=0$ and $k_\perp=\sqrt{k^2_y + k^2_z}$. $\epsilon_{+s'}(\k)$ represents the conduction band and $\epsilon_{-s'}(\k)$ the valence band. There are two branches
$s'=+1$(dashed red curve) which has no nodes and $s'=-1$(dashed blue curve) which involves a nodal circle depicted as the solid blue circle of radius b. The vertical 
arrows show the minimum interband optical transitions possible for the three values of the chemical potential $\mu$ shown $\mu<b$, $\mu=b$ and $\mu>b$. Transitions with 
shorter arrows are Pauli blocked and hence are not possible.} 
\label{fig:Fig1}
\end{figure} 
\be
\label{Dispersion}
\epsilon_{ss'}(\k)=s\sqrt{k^2_{x}+\lp \sqrt{k^2_{y}+k^2_{z}} + s'b\rp^2} = s \epsilon_{s'}(\k)
\ee
where $s'=\pm$ and $s=\pm$. The index $s$ gives conduction ($+1$) and valence ($-1$) band associated with the dispersion curves of the $s'$ band. These dispersion curves 
are highlighted in Fig.~\ref{fig:Fig1}.

The $zz$ component of the dynamical conductivity $\sigma_{zz}(\Omega)$ which can be calculated from the Kubo formula, depends on the spectral density $A_{ss'}(\omega)$ 
of the charge carriers and takes the form,
\bea
\label{OC}
&& \sigma^{IB}_{zz}(\Omega)= \frac{e^2\pi}{\Omega} \int^{+\infty}_{-\infty}\hspace{-0.6cm} d\omega \lb f(\omega)-f(\omega+\Omega)\rb \sum_{ss'} \int\hspace{-0.2cm} \frac{d^3\k}{(2\pi)^3}  \times \nonumber \\ 
&& \lp 1-\frac{k^2_x}{\epsilon^2_{ss'}(\k)}\rp A_{ss'}(\k,\omega)A_{-ss'}(\k,\omega+\Omega),
\eea
for the interband transitions ($\sigma^{IB}_{zz}(\Omega)$) and the intraband part ($\sigma^{D}_{zz}(\Omega)$) is given by the formula 
\bea
\label{Drude}
&& \sigma^{D}_{zz}(\Omega)= \frac{e^2\pi}{\Omega} \int^{+\infty}_{-\infty}\hspace{-0.6cm} d\omega \lb f(\omega)-f(\omega+\Omega)\rb \sum_{ss'} \int\hspace{-0.2cm} \frac{d^3\k}{(2\pi)^3}  \times \nonumber \\ 
&& \frac{k^2_x}{\epsilon^2_{ss'}(\k)} A_{ss'}(\k,\omega)A_{ss'}(\k,\omega+\Omega).
\eea


In terms of the carrier self energy $\Sigma_{ss'}(\omega)$, the spectral functions take the form \cite{Sharapov},
\be
\label{Spectral function}
A_{ss'}(\k,\omega)= \frac{1}{\pi} \frac{|-\Im\Sigma_{ss'}(\omega)|}{\lp \omega -\Re \Sigma_{ss'}(\omega)-\epsilon_{ss'}(\k)\rp^2 + \lp \Im \Sigma_{ss'}(\omega)\rp^2}.
\ee
For simplicity in our calculation we will take the case of residual scattering modeled through a constant imaginary part $|-\Im\Sigma_{ss'}(\omega)|\equiv\Gamma$. We 
will consider two limiting cases for Eq.~(\ref{OC}). The D.C. limit of $\Omega\to 0$ in which instance the thermal factor 
\be 
\lim_{\Omega \to 0}\lf f(\omega)-f(\omega+\Omega)\rf = -\frac{\partial f(\omega)}{\partial \omega},
\ee
where $f(\omega)$ is the Fermi-Dirac thermal distribution function. The zero temperature case for which the integral over $\omega$ becomes limited to the range $\mu$ to 
$\mu-\Omega$ where $\mu$ is the chemical potential and the thermal factor $\lb f(\omega)-f(\omega+\Omega) \rb$ is to be replaced by 1. 

The $T=0$ limit involves an integral over $\omega$ of the form
\be
\label{Basic-Int}
\int^{\mu}_{\mu-\Omega} \frac{d\omega}{\pi^2} \frac{\Gamma}{\lp \omega -\epsilon_{ss'}(\k)\rp^2 +\Gamma^2} \times \frac{\Gamma}{\lp \omega+\Omega-\epsilon_{ss'}(\k)\rp^2 
+\Gamma^2} ,
\ee
which can be done analytically and gives,
\be
\label{IntIB}
\frac{\sigma^{\hspace{-0.1cm}IB}_{\hspace{-0.1cm}zz}\hspace{-0.1cm} \lp\hspace{-0.1cm}T\hspace{-0.1cm}=\hspace{-0.1cm}0,\hspace{-0.07cm}\Omega\rp}{\Gamma}\hspace{-0.1cm}=\hspace{-0.1cm} 
\frac{e^2}{2\pi^3 \hbar^2v_{F}}\hspace{-0.15cm}\sum_{s'} \hspace{-0.2cm}\int^{\hspace{-0.05cm}\infty}_{\hspace{-0.05cm}0}\hspace{-0.2cm}
\frac{d\bk_{x}}{\bOmega}\hspace{-0.2cm}\int^{\hspace{-0.05cm}\infty}_{\hspace{-0.05cm}0}\hspace{-0.35cm} \brho d\brho 
\hspace{-0.1cm}\lp\hspace{-0.15cm}1\hspace{-0.1cm}-\hspace{-0.1cm}\frac{\bk^2_{x}}{\bepsilon^2_{s'}}\hspace{-0.15cm}\rp \hspace{-0.13cm}\Y(\bmu,\bOmega,\bb,\bepsilon_{s'}), 
\ee
where we have introduced polar coordinates for $k_{y},k_{z}$ variables and have divided all variables by the scattering rate which has had the effect of scaling out 
$\Gamma$. Of course it remains in $\widetilde{\Omega}\equiv \frac{\Omega}{\Gamma}, \widetilde{\mu}\equiv \frac{\mu}{\Gamma}$ and $\widetilde{b}\equiv \frac{b}{\Gamma}$ 
while the other variables are dummies of integration. The function $\Y(\bmu,\bOmega,\bb,\bepsilon_{s'})$ has the form 
\bea
\label{Special-function-interband}
&& \Y(\bmu,\bOmega,\bb,\bepsilon_{s'})= \frac{1}{(\bOmega+2\bepsilon_{s'}) \lb 4 + \lp \bOmega + 2\bepsilon_{s'} \rp ^2 \rb} \times \nonumber  \\ 
&& \biggl [\ln \lp \frac{1+(\bmu+\bOmega+\bepsilon_{s'})^2}{1+(\bmu-\bepsilon_{s'})^2} \times \frac{1+(\bmu-\bOmega-\bepsilon_{s'})^2}{1+(\bmu+\bepsilon_{s'})^2}\rp 
+\nonumber  \\
&& (\bOmega+2\bepsilon_{s'}) \{ \arctan(\bmu+\bOmega+\bepsilon_{s'}) + \arctan(\bmu-\bepsilon_{s'}) - \nonumber  \\
&& \arctan(\bmu+\bepsilon_{s'}) - \arctan(\bmu-\bOmega-\bepsilon_{s'}) \} \biggr] +\nonumber  \\
&& \frac{1}{(\bOmega-2\bepsilon_{s'}) \lb 4 + \lp \bOmega - 2\bepsilon_{s'} \rp ^2 \rb} \times \nonumber  \\ 
&& \biggl [\ln \lp \frac{1+(\bmu+\bOmega-\bepsilon_{s'})^2}{1+(\bmu+\bepsilon_{s'})^2} \times \frac{1+(\bmu-\bOmega+\bepsilon_{s'})^2}{1+(\bmu-\bepsilon_{s'})^2}\rp 
+\nonumber  \\
&& (\bOmega-2\bepsilon_{s'}) \{ \arctan(\bmu+\bOmega-\bepsilon_{s'}) + \arctan(\bmu+\bepsilon_{s'}) - \nonumber  \\
&& \arctan(\bmu-\bepsilon_{s'}) - \arctan(\bmu-\bOmega+\bepsilon_{s'}) \} \biggr].
\eea
An equivalent equation can be derived for the intraband case. We obtain,
\be
\label{OC Drude}
\frac{\sigma^{D}_{zz}(T\hspace{-0.1cm}=\hspace{-0.1cm}0,\Omega)}{\Gamma}=\frac{e^2}{2\pi^3 \hbar^2v_{F}}\hspace{-0.15cm} \sum_{s'}\hspace{-0.15cm} 
\int^{\infty}_{0} \hspace{-0.15cm}\frac{d\bk_{x}}{\bOmega} \hspace{-0.2cm}\int^{\infty}_{0}\hspace{-0.4cm} \brho d\brho 
\frac{\bk^2_{x}}{\bepsilon^2_{s'}} \mathfrak{H}(\bmu,\bOmega,\bb,\bepsilon_{s'}), 
\ee
with 
\bea
\label{Special-function-intraband}
&& \mathfrak{H}(\bmu,\bOmega,\bb,\bepsilon_{s'}) = \frac{1}{\bOmega \lb \bOmega^2 + 4\rb} \biggl [ \ln \biggl( \frac{\lp\hspace{-0.1cm}1\hspace{-0.1cm}+\hspace{-0.1cm}(\bmu\hspace{-0.05cm}+\hspace{-0.05cm}\bOmega\hspace{-0.05cm}+\hspace{-0.05cm}\bepsilon_{s'})^2\hspace{-0.1cm}\rp}{1\hspace{-0.1cm}+\hspace{-0.1cm}(\bmu\hspace{-0.05cm}+\hspace{-0.05cm}\bepsilon_{s'})^2} \times \nonumber \\ 
&& \hspace{-0.3cm}\frac{\lp\hspace{-0.1cm}1\hspace{-0.1cm}+\hspace{-0.1cm}(\bmu\hspace{-0.05cm}+\hspace{-0.05cm}\bOmega\hspace{-0.05cm}-\hspace{-0.05cm}\bepsilon_{s'})^2\hspace{-0.1cm}\rp}{1\hspace{-0.1cm}+\hspace{-0.1cm}(\bmu\hspace{-0.05cm}+\hspace{-0.05cm}\bepsilon_{s'})^2} 
\frac{\lp\hspace{-0.1cm}1\hspace{-0.1cm}+\hspace{-0.1cm}(\bmu\hspace{-0.05cm}-\hspace{-0.05cm}\bOmega\hspace{-0.05cm}+\hspace{-0.05cm}\bepsilon_{s'})^2\hspace{-0.1cm}\rp}{1\hspace{-0.1cm}+\hspace{-0.1cm}(\bmu\hspace{-0.05cm}-\hspace{-0.05cm}\bepsilon_{s'})^2}
\frac{\lp\hspace{-0.1cm}1\hspace{-0.1cm}+\hspace{-0.1cm}(\bmu\hspace{-0.05cm}-\hspace{-0.05cm}\bOmega\hspace{-0.05cm}-\hspace{-0.05cm}\bepsilon_{s'})^2\hspace{-0.1cm}\rp}{1\hspace{-0.1cm}+\hspace{-0.1cm}(\bmu\hspace{-0.05cm}-\hspace{-0.05cm}\bepsilon_{s'})^2}\hspace{-0.05cm} \biggr)\hspace{-0.05cm}+\hspace{-0.05cm} \nonumber \\
&&\bOmega \{ \arctan(\bmu\hspace{-0.05cm}+\hspace{-0.05cm}\bOmega\hspace{-0.05cm}+\hspace{-0.05cm}\bepsilon_{s'})\hspace{-0.1cm}+\hspace{-0.1cm}\arctan(\bmu\hspace{-0.05cm}+\hspace{-0.05cm}\bOmega\hspace{-0.05cm}-\hspace{-0.05cm}\bepsilon_{s'})\hspace{-0.05cm}- \hspace{-0.05cm}\nonumber \\
&& \arctan(\bmu\hspace{-0.05cm}-\hspace{-0.05cm}\bOmega\hspace{-0.05cm}+\hspace{-0.05cm}\bepsilon_{s'})\hspace{-0.1cm}-\hspace{-0.1cm}\arctan(\bmu\hspace{-0.05cm}-\hspace{-0.05cm}\bOmega\hspace{-0.05cm}-\hspace{-0.05cm}\bepsilon_{s'}) \} \biggr].
\eea
We have checked that these equations properly reduce to the clean limit forms for $\Gamma\rightarrow 0$ used in the work of Ref.[\onlinecite{Carbotte2}] with which we 
will compare when appropriate.

\section{Results for finite frequencies}
\label{sec:III}

\begin{figure}
\includegraphics[width=2.5in,height=3.4in, angle=270]{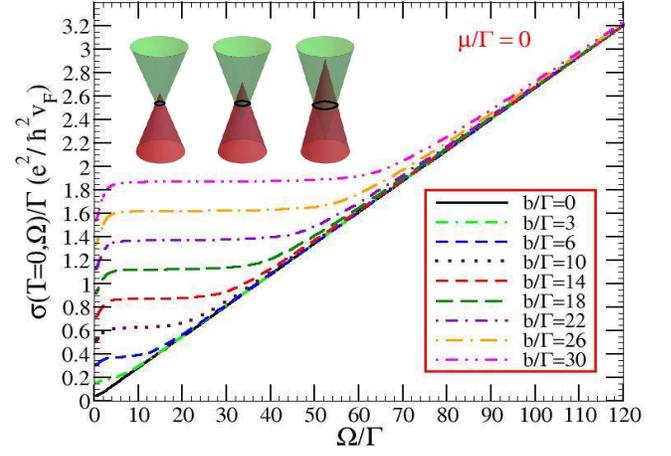} 
\caption{(Color online) The conductivity $\sigma(T=0,\Omega)$ normalized by the scattering rate $\Gamma$ in units of $\frac{e^2}{\hbar^2 v_F}$ as a function of normalized
photon energy $\Omega/\Gamma$. Various values of $b/\Gamma$ are shown with b the loop node radius in energy units. The chemical potential $\mu$ is set equal to zero. 
Except for the smaller values of $b/\Gamma$ the 2D graphene like constant background conductivity is well developed for photon energies $\Omega$ below $2b$. For photon 
energies above $2b$ the linear law of the 3D Dirac point node case is recovered. At small values of $\Omega$ there is a downturn of the constant background to its 
minimum D.C. value at $\Omega=0$.}
\label{fig:Fig2}
\end{figure}

In Fig.~\ref{fig:Fig2} we show our results for the zero temperature dynamic optical response $\sigma(T=0,\Omega)$ as a function of photon energy $\Omega$ both quantities
normalized by the optical scattering rate $\Gamma$ i.e. $\sigma(T=0,\Omega)/\Gamma$ vs. $\Omega/\Gamma$. In these reduced variables $\Gamma$ has dropped out and we have 
a single set of curves which apply for any value of $\Gamma$. In fact we have a family of curves defined by two external parameters, the normalized chemical potential 
$\mu/\Gamma\equiv\bmu$ and nodal loop parameter $b/\Gamma\equiv\bb$. Here $\bmu=0$ represents charge neutrality and we consider nine values of $b/\Gamma$. Starting from 
the top curve we see a graphene like 2D constant background extending from a bit above of $\Omega=0$ all the way to $\Omega=2b$ at which point it shows a transition to a 
linear in $\Omega/\Gamma$ behavior characteristic of a 3D point node Dirac material. The height of this plateau referred to as the interband background \cite{Gusynin} 
and denoted by $\sigma^{BG}$ agrees (after a correction of a dropped factor of 2) with 
the value obtained in Ref.[\onlinecite{Carbotte2}] where the clean limit $\Gamma\rightarrow 0$ was considered. At $\Omega/\Gamma\rightarrow 0$ however our new results show 
a bend downward to connect with the D.C. value of the conductivity as we will discuss shortly. As the value of the nodal loop parameter is reduced towards zero the 
frequency range over which a plateau is well defined shrinks. It is still seen in the dashed blue curve for $b/\Gamma=6$ but below this value we see a clear evolution 
from a constant value to linear like behavior characteristic of a 3D Dirac point node system as we will elaborate upon in Fig.~\ref{fig:Fig4}. First we return to the 
$\Omega=0$ limit. As we will see later (Eq.~\ref{DCmu0}) we get a particularly simple and important result
\be
\label{Sigma DC}
\sigma^{DC}= \frac{e^2}{\hbar^2 v_{F}} \frac{1}{2\pi^2} \sqrt{\Gamma^2+b^2}.
\ee

In the limit $b=0$ which here corresponds to a 3D point node Dirac model we get the known answer obtained in Ref.[\onlinecite{Nicol}], namely the minimum conductivity
$\sigma^{DC}$ is not universal but rather depends linearly on optical scattering rate ( tewice the quasiparticle rate $\Gamma$). This is a very different result from 
that obtained for graphene \cite{Sharapov} for which $\sigma^{DC}$ is simply a number $\frac{4e^2}{\pi h}$ in the same constant $\Gamma$ approximation used here and 
referred to as the universal interband background ($\sigma^{BG}$). We see however from Eq.~\ref{Sigma DC} that in the nodal loop 
semimetal $\Gamma$ drops out of $\sigma^{DC}$ if $b>> \Gamma$ and in that case we get no dependence on scattering rate $\Gamma$ so $\sigma^{DC}$ is again universal equal 
to $\frac{e^2}{\hbar^2 v_{F}}\frac{b}{2\pi^2}$. While this minimum conductivity does not depend on $\Gamma$ it is linear in $b$ and inversely dependent on $v_{F}$ which 
are material dependent properties. In graphene no such material parameters arise \cite{Sharapov}. A universal conductivity \cite{Lee} is also part of d-wave \cite{Jiang} 
superconductivity theory. It arises when a gap which can have complex symmetry \cite{Donovan, Donovan1} nevertheless goes through a zero \cite{Yip, Schachinger} on the 
Fermi surface. It does not arise in an s-wave superconductor even if there is some anisotropy such as in Al\cite{Leung} but with no zero.
We note that the magnitude of minimum conductivity depends on the model used for the disorder as discussed in the review of Evers and Mirlin [\onlinecite{Mirlin}] for the 
specific case of graphene. Here only the simplest model of constant scattering was used.

\begin{figure}
\includegraphics[width=2.5in,height=3.4in, angle=270]{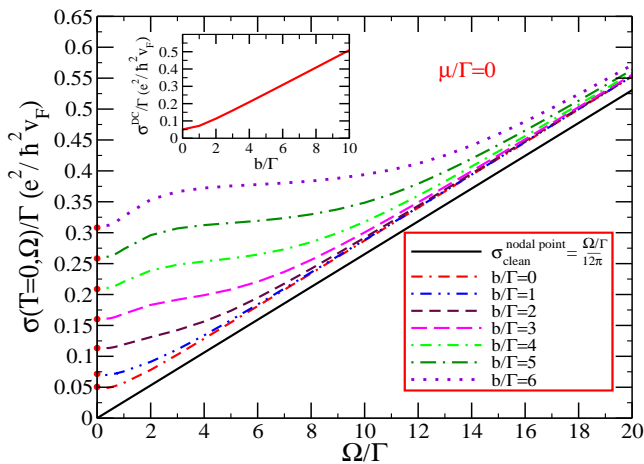} 
\caption{(Color online) The conductivity $\sigma(T=0,\Omega)$ normalized by the scattering rate $\Gamma$ in units of $\frac{e^2}{\hbar^2 v_F}$ as a function of normalized
photon energy $\Omega/\Gamma$. We show the transition from a 2D graphene like constant background for $\Omega<2b$ (dotted violet curve) to the linear in $\Omega$ 
behavior of a 3D Dirac point node as $b/\Gamma$ is reduced from a value of six to zero when the double-dashed-dotted red curve applies. This last curve still differs 
from the solid black curve which applies in the clean limit and is a straight line of slope $\frac{1}{12\pi}$ in the our units. The D.C. limit of these curves 
($\sigma^{DC}(T=0, \mu=0)$) is shown with heavy red points on the $\Omega/\Gamma=0$ axis and is never zero when some scattering is included. The inset shows how it 
evolves as a function of $b/\Gamma$.}
\label{fig:Fig3}
\end{figure}

The approach to the minimum D.C. conductivity and more generally the transition from 2D graphene like behavior for the A.C. conductivity to 3D point node Dirac behavior
as $b$ gets small is further elaborated upon in Fig.~\ref{fig:Fig3}. Again we show the photon energy dependence of the zero temperature conductivity $\sigma(T=0,\Omega)$ 
both normalized by $\Gamma$ so as to get universal curves but here we consider only values of $b$ less than $b/\Gamma=6$ (dotted violet curve). While this curve still 
shows a clear tendency to flatten out in the region around $\Omega/\Gamma\approx 5$, the other results do not. By $b/\Gamma=1$ the curve for $\sigma(T=0,\Omega)/\Gamma$ 
vs $\Omega/\Gamma$ is convex upward and no trace of a plateau remains. The dot-double-dashed red curve for $b=0$ reproduces the results of  Ref.[\onlinecite{Nicol}] for 
a 3D Dirac point node semimetal. We have also placed on the same graph their results (solid black curve) for the clean limit. A straight line of slope $\frac{1}{12\pi}$ 
applies in this case . In the inset to Fig.~\ref{fig:Fig3} we show the evolution of minimum D.C. conductivity $\sigma^{DC}$ normalized to $\Gamma$ as a function of 
$b/\Gamma$ from Eq.~\ref{Sigma DC}, which in our units is $\frac{1}{2\pi^2}\sqrt{1+(b/\Gamma)^2}$. We see that it rapidly goes from a constant $\frac{1}{2\pi^2}$ to 
$\frac{b/\Gamma}{2\pi^2}$ linear in $b$ and inversely proportional to $\Gamma$. These results are plotted on the $\Omega=0$ axis of the main frame as heavy red dots.

\begin{figure}
\includegraphics[width=2.5in,height=3.4in, angle=270]{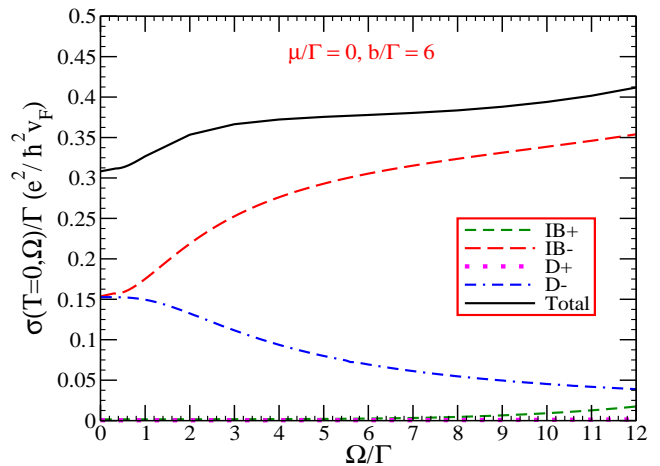} 
\caption{(Color online) The conductivity $\sigma(T=0, \Omega)$ normalized by the scattering rate $\Gamma$ in units of $\frac{e^2}{\hbar^2 v_F}$ as a function of normalized
photon energy $\Omega/\Gamma$ for the case $\mu/\Gamma=0$ and $b/\Gamma=6$ (solid black line). The other curves shows the decomposition of the conductivity in terms
of the intra (D) and inter (IB) band optical transitions and the two bands $s'=\pm$. The $s'=+1$ band is gaped by b and does not contribute much. The $s'=-1$ band in 
contrast shows that at $\Omega =0$ (D.C. limit) both intra (Drude) and interband contributions are equal. As $\Omega/\Gamma$ is increased IB$_{-}$ increases while D$_{-}$ 
decreases as we expect.}
\label{fig:Fig4}
\end{figure}

It is important to understand that when residual scattering (finite $\Gamma$) is introduced both interband and intraband optical transition contribute to the minimum 
conductivity Eq.~\ref{Sigma DC} as they do to finite photon energy properties. This is illustrated in Fig.~\ref{fig:Fig4} where we treat the specific case 
$\mu/\Gamma=0$, $b/\Gamma=6$ for definiteness. We have decomposed the contributions to $\sigma(T=0,\Omega)$ into four terms. The dashed green line is the interband (IB) 
contribution of the $s'=+1$ band and the dashed red line is for $s'=-1$, the dotted magenta line is the intraband (D) contribution with $s'=+1$ and the double-dashed-dotted 
blue line for $s'=-1$. It is clear and expected that the $s'=+1$ band contributes little to the conductivity in the region $\Omega<2b$ because it is gaped as can be seen 
in Fig.~\ref{fig:Fig1}. In fact if we had not included some scattering ($\Gamma=0$) these contributions would be identically zero in the clean limit. Even for 
$\Gamma\ne 0$ it is the $s'=-1$ that gives almost the entire contribution from interband (dashed red curve) and intraband (double-dashed-dotted blue curve) optical 
transitions. Note that at $\Omega=0$ (D.C. limit) both give exactly the same contribution. This is entirely due to the presence of finite $\Gamma$. In the clean limit 
there is no intraband conductivity because the chemical potential $\mu=0$ in our example and the entire D.C. conductivity comes from the interband transitions. Thus 
scattering has a profound effect on these results.

\begin{figure}
\includegraphics[width=2.5in,height=3.4in, angle=270]{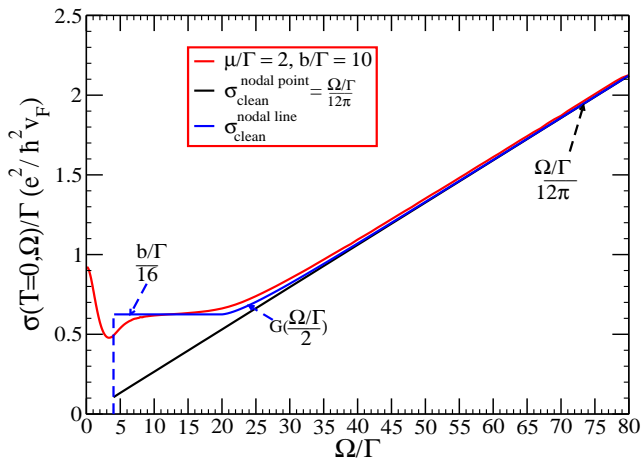} 
\caption{(Color online)The conductivity $\sigma(T=0, \Omega)$ normalized by the scattering rate $\Gamma$ in units of $\frac{e^2}{\hbar^2 v_F}$ as a function of normalized
photon energy $\Omega/\Gamma$. A case of chemical potential $\mu/\Gamma=2$ and $b/\Gamma=10$ (solid red curve) is compared with the clean limit results obtained in 
Ref. [\onlinecite{Carbotte2}] for a loop node semimetal (solid blue curve) as well as with the 3D Dirac point node model (solid black curve) in Ref.[\onlinecite{Nicol}]. 
At the higher values of $\Omega$ shown all these curves merge and define a straight line of slope $\frac{1}{12\pi}$ in our units. This straight line continues in the 
point node case until $\Omega=2\mu$ where it drops to zero. The lost optical spectral weight is transferred to a Dirac delta function at $\Omega=0$ (not seen here). For 
the nodal loop there is a short region in photon energy where the straight line of slope $\frac{1}{12\pi}$ changes to a constant of height $\frac{b/\Gamma}{16}$ at 
$\Omega=2b$ and is again cut off at $\Omega=2\mu$. The solid red line is different but still shows a plateau like region below $\Omega=2b$ but there is no sharp cutoff
at $\Omega=2\mu$ for two reasons. First the optical weight transferred to the intraband transitions is now broaden into a Drude part of width $2\Gamma$ and the interband
background is also smeared out by the residual scattering.}
\label{fig:Fig5}
\end{figure}

In Fig.~\ref{fig:Fig5} we elaborate further on the relationship of our results to the clean limit results of Ref.[\onlinecite{Nicol, Carbotte2}]. What is shown as the 
solid red curve is $\sigma(T=0,\Omega)$ normalized with $\Gamma$ in units of $\frac{e^2}{\hbar^2 v_{F}}$ as a function of $\Omega/\Gamma$ for a case $\mu/\Gamma=2$ and 
$b/\Gamma=10$. In this example a Drude like peak is clearly seen in the vicinity of $\Omega/\Gamma\approxeq0$. This contribution would exists even in the clean limit but 
would take the form of a Dirac delta function at $\Omega=0$. For comparison with the clean limit we also show as solid blue line the clean limit result of reference 
Ref.[\onlinecite{Carbotte2}] corrected for a missing factor of 2. We have,
\be
\label{SigmaIB}
\frac{\sigma^{IB}(\Omega)}{\Gamma} = \frac{e^2 \pi}{\hbar^2 v_{F}} \frac{1}{2\Omega \Gamma} I(\frac{\Omega}{2}),
\ee
with 
\bea
\label{IOmega}
&& I(\omega)= \frac{\omega b}{4\pi} ~~~~~  \omega<b, \nonumber \\
&& =\frac{1}{2\pi^2} \lb b|\omega|\arctan\frac{b}{\sqrt{\omega^2-b^2}} + \frac{\sqrt{\omega^2-b^2}}{3|\omega|} \lp 4\omega^2-b^2\rp \rb  \nonumber\\
&& \hspace{6cm}\omega>b.
\eea
This is plotted as blue solid line (labeled as $G(\frac{\Omega/\Gamma}{2})$) which is cut off below $2\mu$ because interband optical transition are not possible below 
this photon energy because of Pauli blocking (see Fig.~\ref{fig:Fig1}). The optical spectral weight lost is of course transferred to a delta function at $\Omega=0$ not 
shown on the figure. Having understood that at finite $\Gamma$ this spectral weight is distributed into a Drude like form around $\Omega=0$ we see that the clean limit 
results agree very well with our new results for the finite $\Gamma$ case (solid red curve). We have also placed for additional comparison the results of 
Ref.[\onlinecite{Nicol}] for point node Dirac as a solid black curve which is a straight line of slope $\frac{1}{12\pi}$. It too is to be cut off at $\Omega=2\mu$. 
While it matches well with the other two curves in the large $\Omega$ region it is very different from the nodal loop results below $\Omega=2 b$.

\begin{figure}
\includegraphics[width=2.5in,height=3.4in, angle=270]{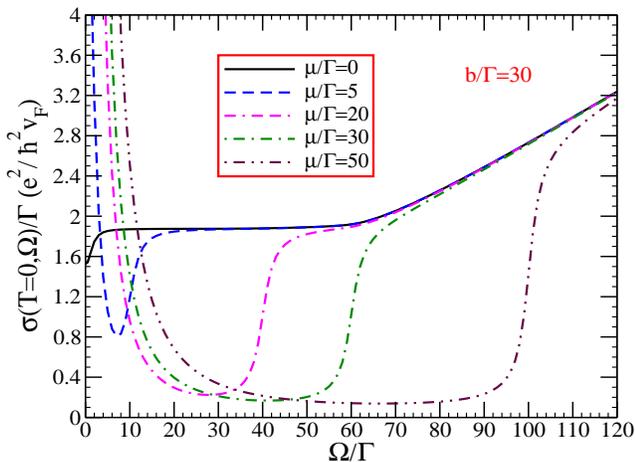} 
\caption{(Color online)The conductivity $\sigma(T=0, \Omega)$ normalized by the scattering rate $\Gamma$ in units of $\frac{e^2}{\hbar^2 v_F}$ as a function of normalized
photon energy $\Omega/\Gamma$. In all cases $b/\Gamma$ is set to be 30 while five different values of chemical potential are considered ranging from $\mu/\Gamma=0$ to 
$\mu/\Gamma=50$ which is almost twice the value of $b/\Gamma$ chosen. Except for the $\mu/\Gamma=0$ case (solid black line) the curves all show a Drude like peaks at 
small values of $\Omega/\Gamma$. These peaks increases with the value of $\mu/\Gamma$ as a result of increasing optical spectral weight transfer from the interband to 
intraband optical transitions. When $\mu/\Gamma$ is small compared with $b/\Gamma$ the 2D graphene like constant background is recovered as  $\Omega/\Gamma$ increases 
while for the last curve with $\mu>b$ the recovery is to the 3D Dirac point node linear in $\Omega$ law.}
\label{fig:Fig6}
\end{figure}

In Fig.~\ref{fig:Fig6} we present additional results for the case of finite chemical potential. Here $b/\Gamma$ is fixed at 30, and five values of $\mu/\Gamma$ are 
considered. The solid black line is for $\mu=0$ and has already been presented in Fig.~\ref{fig:Fig2} as dashed-double-dotted magenta line. It is repeated here for 
comparison. The finite $\mu$ curves all shows a large Drude like peaks at small photon energies. Beyond this Drude, all curves show a depression in conductivity 
before rising up again to meet the $\mu=0$ curve at higher energies. The dashed blue curve for $\mu/\Gamma=5$ has only a small dip and has recovered to its $\mu=0$ 
value by $\Omega/\Gamma=20$; which is twice the value of $2\mu$, the low energy cut off on interband transition that would apply to the clean limit. Here this cutoff 
is no longer sharp because of the smearing caused by disorder (finite $\Gamma$). When the chemical potential is increased to $\mu/\Gamma=20$ (double-dashed-dotted magenta 
curve) the conductivity displays a much more pronounced dip beyond the Drude region of photon energy. In the clean limit we would have had a large delta function 
contribution at $\Omega=0$ than a complete zero up to $2\mu/\Gamma=40$ after which it would have recovered its plateau value. Disorder smearing has partially filled in 
the gap region and has also smeared the edge at $2\mu$. At still higher values of $\mu$ the same behavior is observed but at large $\Omega/\Gamma$ the conductivity goes 
back to the linear law of 3D point node Dirac.

\begin{figure}
\includegraphics[width=2.5in,height=3.4in, angle=270]{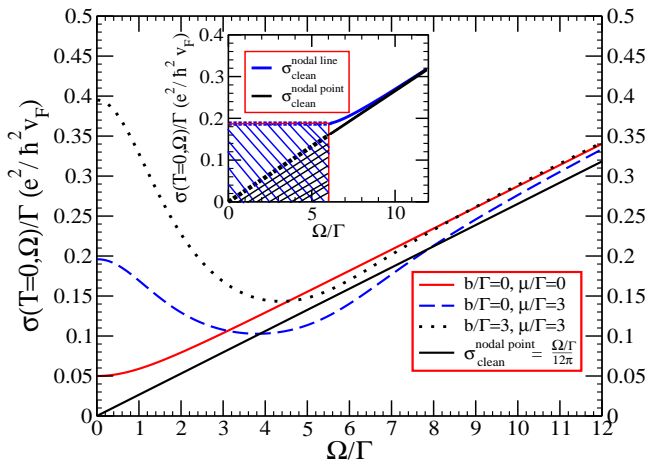} 
\caption{(Color online)The optical conductivity $\sigma(T=0, \Omega)$ normalized by the scattering rate $\Gamma$ in units of $\frac{e^2}{\hbar^2 v_F}$ as a function of 
normalized photon energy $\Omega/\Gamma$. The case of $\mu/\Gamma=b/\Gamma=3$ (dotted black curve) is compared with the case $\mu/\Gamma=3$ but $b/\Gamma=0$ (dashed 
blue curve). Also shown as solid red line is the case $\mu = b = 0$ and solid black the clean limit. The inset shows results for the clean limit for $\mu=b$ in loop 
nodes (solid blue) and point node (solid black). The shaded region below $\Omega=2b$ is entirely transferred to the intraband transition. The ratio of the optical 
spectral weight transferred of loop and point node $\frac{OSW_{loop}}{OSW_{point}}$ is equal to $\frac{3\pi}{4}$.}
\label{fig:Fig7}
\end{figure}

One interesting special case with some additional commentary is the case $\mu=b$. In the clean limit this corresponds to the transfer of the entire optical spectral weight 
from the flat 2D graphene like plateau in the interband conductivity to the intraband delta function at $\Omega=0$. This leave a region above $\Omega=2b$ which would
deviate very little from the straight line of slope $\frac{1}{12\pi}$ of 3D point node Dirac. Nevertheless the two cases can be differentiated from each other when 
$\Gamma\ne 0$ as is shown in Fig.~\ref{fig:Fig7}. The black dotted line is the result for $b/\Gamma=3$ with $\mu/\Gamma=3$ as well. We see that the disorder smearing has 
broaden the intraband transition contribution to such an extent that no Pauli blocking gap is seen at $2\mu/\Gamma=6$. This is also true for the dashed blue curve obtained 
when $b$ is set to zero. This curve does differ from the black dotted curve when $\Gamma$ is included. The main reason for this is that the optical spectral weight under 
the Drude is very different in the two cases as is illustrated in the inset of the figure. The shaded region shows the lost optical spectral weight (OSW) in the 
interband background that has been transferred to the intraband. In the clean limit we have $OSW_{loop}=\frac{e^2}{\hbar^2 v_{F}}\frac{b^2}{8}$ for the nodal loop and 
$OSW_{point}=\frac{e^2}{\hbar^2 v_{F}}\frac{b^2}{6\pi}$ for the 3D point node Dirac. The ratio of the loop to point node is $\frac{3\pi}{4}$. Thus there is more than a 
factor of 2 difference between these two quantities and this leads to the striking differences between dotted black and dashed blue curves of the main frame of 
Fig.~\ref{fig:Fig7} in the region of photon energy below twice the value of the chemical potential. Above $\Omega/\Gamma \approxeq 10$ both curves are the same and are 
not very different from the solid red curve for $b=0$ (3D point node Dirac) including finite $\Gamma$ and from the solid black curve which is the clean limit version of 
the solid red curve included for comparison.

\section{D.C. Transport}
\label{sec:IV}

\begin{figure}
\includegraphics[width=2.5in,height=3.4in, angle=270]{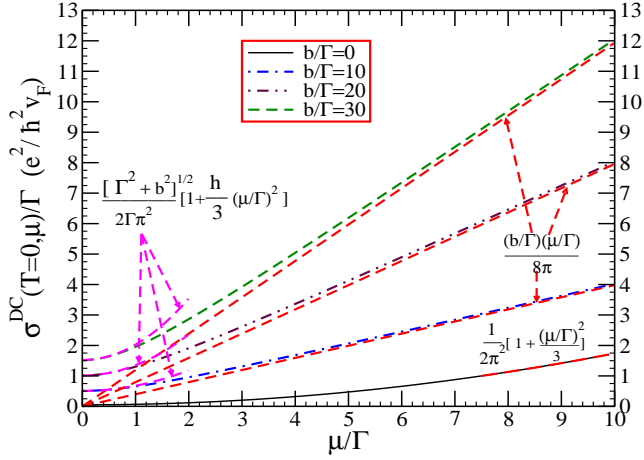} 
\caption{(Color online) The D.C. conductivity $\sigma^{DC}(T=0,\mu)$ normalized by the scattering rate $\Gamma$ in units of $\frac{e^2}{\hbar^2 v_F}$ as a 
function of chemical potential $\mu/\Gamma$ for various values of $b/\Gamma$. The solid black curve for $b/\Gamma=0$ represents the case of 3D Dirac while all others are 
for the nodal loop case. In all these cases the asymptotic value at small $\mu$ and large $\mu$ are indicated by dashed magenta and red lines respectively. At small 
$\mu/\Gamma$ a quadratic in $\mu$ law holds while at large $\mu/\Gamma$ we have the linear in $\mu$ dependence found in Ref.[\onlinecite{Carbotte2}] which treated the 
clean limit. In the present context this limit means both $\mu/\Gamma$ and $b/\Gamma$ are large.}
\label{fig:Fig8}
\end{figure}

\begin{figure}
\includegraphics[width=2.5in,height=3.4in, angle=270]{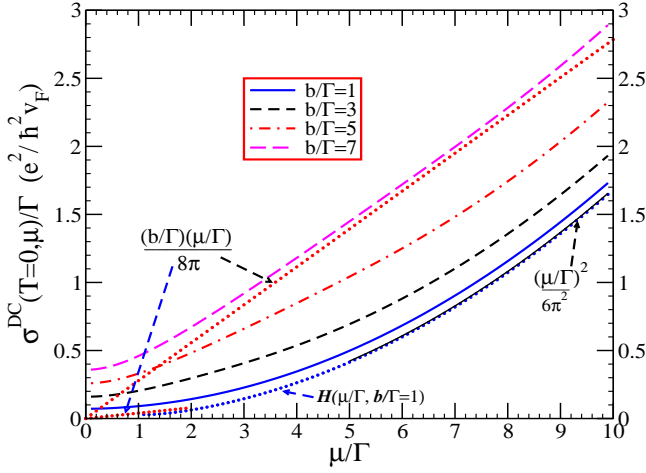} 
\caption{(Color online) The DC conductivity $\sigma^{DC}(T=0,\mu)$ normalized by the scattering rate $\Gamma$ in units of $\frac{e^2}{\hbar^2 v_F}$ as a 
function of normalized chemical potential $\mu/\Gamma$ for $4$ values of $b/\Gamma$ namely $b/\Gamma=1$ (solid blue), $b/\Gamma=3$ (dashed black), $b/\Gamma=5$ (dot 
dashed red) and $b/\Gamma=7$ (dashed magenta). The dotted blue curve shows results in the clean limit for $b/\Gamma=1$ which is described by 
$\mathcal{H}(\mu/\Gamma,b/\Gamma)$ defined in Eq.~(\ref{CurlH}). The solid black curve shows the quadratic in $\mu/\Gamma$ behavior and matches with the blue dotted 
curve for values of $\mu/\Gamma>b/\Gamma $. The dotted red curve is for $b/\Gamma=7$ but only the linear part of $\mathcal{H}(\mu/\Gamma,b/\Gamma)$ is shown. 
Deviations of the dashed magenta curve from this linear law at small values of $\mu/\Gamma$ are due to a finite scattering rate while at large $\mu/\Gamma$ we start 
seeing deviations due to the evolution of $\mathcal{H}(\mu/\Gamma,b/\Gamma)$ from linear in $\mu/\Gamma$ (2D graphene like regime) to a quadratic law (3D Dirac point 
node regime).}
\label{fig:Fig9}
\end{figure}

The D.C. limit of Eq.~(\ref{OC}) gives
\bea
\label{DCIB}
&& \sigma^{IB}_{zz}(\Omega=0)= \frac{2e^2\pi}{\hbar^2 v_{F}}\hspace{-0.1cm} \int^{+\infty}_{-\infty}\hspace{-0.5cm}  d\omega \lp -\frac{\partial f(\omega)}{\partial \omega}\rp \sum_{s'} \int \hspace{-0.2cm} \frac{d^3\k}{(2\pi)^3} \times \nonumber \\ 
&& \hspace{-0.2cm} \lp 1\hspace{-0.1cm}-\hspace{-0.1cm} \frac{k^2_x}{\epsilon^2_{s'}(\k)}\rp \hspace{-0.1cm} \frac{\Gamma^2}{\pi^2} \hspace{-0.1cm} \lp \frac{1}{\Gamma^2\hspace{-0.1cm}+\hspace{-0.1cm}(\omega\hspace{-0.05cm}-\hspace{-0.05cm}\epsilon_{s'})^2} \times 
\frac{1}{\Gamma^2\hspace{-0.1cm}+\hspace{-0.1cm}(\omega\hspace{-0.05cm}+\hspace{-0.05cm}\epsilon_{s'})^2}\rp 
\eea
for the interband contribution and 
\bea
\label{DCD}
&& \sigma^{D}_{zz}(\Omega=0)= \frac{e^2\pi}{\hbar^2 v_{F}} \int^{+\infty}_{-\infty}\hspace{-0.5cm} d\omega \lp -\frac{\partial f(\omega)}{\partial \omega}\rp  \sum_{s'} \int \frac{d^3\k}{(2\pi)^3} \times \nonumber\\ 
&& \hspace{-0.2cm}\frac{k^2_x}{\epsilon^2_{s'}(\k)} \frac{\Gamma^2}{\pi^2} \biggl [ \lp\frac{1}{\Gamma^2\hspace{-0.1cm}+\hspace{-0.1cm}(\omega\hspace{-0.05cm}-\hspace{-0.05cm}\epsilon_{s'})^2}\rp^2 \hspace{-0.1cm}+\hspace{-0.1cm} \lp \frac{1}{\Gamma^2\hspace{-0.1cm}+\hspace{-0.1cm}(\omega\hspace{-0.05cm}+\hspace{-0.05cm}\epsilon_{s'})^2}\rp^2 \biggr ],
\eea
for the intraband or Drude contribution. At zero temperature the thermal factor $-\frac{\partial f(\omega)}{\partial \omega}$ reduces to a delta function of the form 
$\delta(\omega-\mu)$ which pins $\omega$ to be at the chemical potential. Eq.~(\ref{DCIB}) and (\ref{DCD}) then reduce to 2D integrals over $k_{x}$ and $\rho$ which are 
the integration variables introduced in Eq.~(\ref{IntIB}). We get,
\bea
\label{DCIBFinal}
&& \sigma^{IB}_{zz}(\Omega=0)= \frac{2e^2\pi}{\hbar^2 v_{F}} \sum_{s'}\int^{+\infty}_{0} \hspace{-0.3cm}\frac{\rho d\rho}{(2\pi)^2} \int^{+\infty}_{-\infty} \hspace{-0.6cm}dk_{x} 
\frac{\Gamma^2}{\pi^2} \times \nonumber\\ 
&& \hspace{-0.25cm}\lp 1-\frac{k^2_x}{\epsilon^2_{s'}(\k)}\rp  \lp \frac{1}{\Gamma^2\hspace{-0.1cm}+\hspace{-0.1cm}(\mu\hspace{-0.05cm}-\hspace{-0.05cm}\epsilon_{s'})^2} \times  \frac{1}{\Gamma^2\hspace{-0.1cm}+\hspace{-0.1cm}(\mu\hspace{-0.05cm}+\hspace{-0.05cm}\epsilon_{s'})^2}\rp,
\eea
and 
\bea
\label{DCDFinal}
&& \sigma^{D}_{zz}(\Omega=0)= \frac{2e^2\pi}{\hbar^2 v_{F}} \sum_{s'}\int^{+\infty}_{0}\hspace{-0.3cm} \frac{\rho d\rho}{(2\pi)^2} \int^{+\infty}_{-\infty}\hspace{-0.6cm} dk_{x} 
\frac{\Gamma^2}{\pi^2} \times \nonumber\\ 
&& \hspace{-0.2cm}\frac{k^2_x}{\epsilon^2_{s'}(\k)}  \lb \lp\frac{1}{\Gamma^2\hspace{-0.1cm}+\hspace{-0.1cm}(\mu\hspace{-0.05cm}-\hspace{-0.05cm}\epsilon_{s'})^2}\rp^2 \hspace{-0.2cm}+\hspace{-0.1cm} \lp \frac{1}{\Gamma^2\hspace{-0.1cm}+\hspace{-0.1cm}(\mu\hspace{-0.05cm}+\hspace{-0.05cm}\epsilon_{s'})^2}\rp^2 \rb. 
\eea
Results for $\sigma^{DC}(T=0, \mu)\equiv \sigma^{IB}_{zz}({\Omega=0}) + \sigma^{D}_{zz}(\Omega=0)$ are given in Fig.~\ref{fig:Fig8} where we plot $\sigma^{DC}/\Gamma$ 
in units of $\frac{e^2}{\hbar^2 v_{F}}$ as a function of $\mu/\Gamma$ for various values of $b/\Gamma$ namely $b/\Gamma=0$ solid black curve, $b/\Gamma=10$ 
dashed-dotted blue, $b/\Gamma=20$ double-dotted-dashed brown and $b/\Gamma=30$ dashed green. The solid black line is known from the 3D point node case \cite{Nicol} and is 
\be
\frac{\sigma^{DC}(T=0, \mu)}{\Gamma}= \frac{e^2}{\hbar^2 v_{F}} \frac{1}{2\pi^2} \lb 1 + \frac{1}{3} \frac{\mu^2}{\Gamma^2}\rb.
\ee
The limit of zero chemical potential gives
\be
\label{DCmu0}
\frac{\sigma^{DC}(T=0, \mu=0)}{\Gamma}= \frac{e^2}{\hbar^2 v_{F}} \lb\frac{\sqrt{\Gamma^2+b^2}}{2\pi^2\Gamma}\rb.
\ee
In this case the sum of Eq.~(\ref{DCIBFinal}) and (\ref{DCDFinal}) is particularly simple and the terms $\lp 1-\frac{k^2_x}{\epsilon^2_{s'}(\k)}\rp$ in 
Eq.~(\ref{DCIBFinal}) and $\frac{k^2_x}{\epsilon^2_{s'}(\k)}$ in Eq.~(\ref{DCDFinal}) add to give 1. We get,
\be
\label{DCTot}
\hspace{-0.28cm}\sigma^{DC}(T\hspace{-0.1cm}=\hspace{-0.1cm}0,\mu\hspace{-0.1cm}=\hspace{-0.1cm}0)\hspace{-0.1cm}= \hspace{-0.1cm}\frac{2e^2\pi}{\hbar^2 v_{F}}\hspace{-0.15cm} \sum_{s'}\hspace{-0.2cm}\int^{\hspace{-0.05cm}\infty}_{\hspace{-0.05cm}0}\hspace{-0.3cm}\frac{\rho d\rho}{(2\pi)^2} \hspace{-0.2cm}\int^{\hspace{-0.05cm}\infty}_{\hspace{-0.1cm}0}\hspace{-0.45cm}dk_{x}
\frac{\Gamma^2}{\pi^2}\hspace{-0.05cm} \frac{1}{\lp\Gamma^2+\epsilon^2_{s'}\rp^2},
\ee
which can be done analytically and gives Eq.~(\ref{DCmu0}). The implications of Eq.~(\ref{DCmu0}) were discussed in the previous section. Here we note that except for 
the $b/\Gamma=0$ curve $\sigma^{DC}(\mu=0)$ increases linearly with b in Fig.~\ref{fig:Fig8} because $b/\Gamma>>1$. We shall also show below that for finite $\mu$ but 
$\mu/\Gamma<1$, $\sigma^{DC}(\mu)$ increases as $\mu^2/\Gamma^2$ out of its $\mu=0$ value. This behavior is indicated by a dashed magenta curve which agrees with our full 
numerical results at small $\mu$ but deviation occur as $\mu$ increases. In fact, for the finite $b$ curves there is a gradual evolution to a linear in $\mu$ dependence 
of $\sigma^{DC}(\mu)$ which is characteristic of 2D graphene like behavior as obtained in Ref.[\onlinecite{Carbotte2}] for the clean limit. Here the clean limit 
corresponds to $\mu/\Gamma$ and $b/\Gamma$ large ($\Gamma\rightarrow 0$). For $\mu<b$, the clean limit result is 
\be
\frac{\sigma^{DC}(T=0, \mu)}{\Gamma}= \frac{e^2}{\hbar^2 v_{F}} \frac{(b/\Gamma)(\mu/\Gamma)}{8\pi}.
\ee
which is shown in Fig.~\ref{fig:Fig8} as the dashed red lines that go through the origin. The main difference between clean limit results and those that included carrier 
scattering are at small values of $\mu/\Gamma$ (near charge neutrality). Including a finite $\Gamma$ changes the dependence of the D.C. conductivity from linear in 
$\mu/\Gamma$ to a quadratic dependence in the region $\mu/\Gamma<1$. By contrast the solid black curve for $b=0$ follows a quadratic law $\mu^2/\Gamma^2$ over the entire 
range shown. The evolution from quadratic to linear behavior is studied in more detail in Fig.~\ref{fig:Fig9} where small values of $b$ are considered namely $b=1$ 
(solid blue), $b=3$ (dashed black), $b=5$ (dashed-dotted red) and  $b=7$ (dashed magenta) curves. Also shown on the figure for comparison are clean limit results (dotted 
blue) for $b=1$ and (dotted red) for $b=7$. What is plotted for the dotted blue line are the result for the clean limit from Ref.[\onlinecite{Carbotte2}] 
(again corrected for a missed factor of 2). The relevant function is,
\be
\frac{\sigma^{DC}(T=0,\mu)}{\Gamma} = \frac{e^2}{\hbar^2 v_{F}}\mathcal{H}(\frac{\mu}{\Gamma},\frac{b}{\Gamma}),
\ee
with
\bea
\label{CurlH}
&& \mathcal{H}(\frac{\mu}{\Gamma},\frac{b}{\Gamma}) = \frac{(\mu/\Gamma)(b/\Gamma)}{8\pi}         ~~~~~~~~~~~\mu<b, \nonumber \\
&& =\frac{1}{2\pi^2} \biggl[ \frac{1}{2} (\frac{\mu}{\Gamma})(\frac{b}{\Gamma}) \arctan \lp \frac{b/\Gamma}{\sqrt{(\mu/\Gamma)^2-(b/\Gamma)^2}}\rp + \nonumber \\
&& \hspace{-0.3cm}\frac{1}{3} \lp \frac{(\mu/\Gamma)^2-\frac{3}{2}(b/\Gamma)^2}{\mu/\Gamma}\rp \sqrt{(\mu/\Gamma)^2-(b/\Gamma)^2}\biggr] \mu>b. 
\eea
For $\mu>>b$, $\mathcal{H}(\frac{\mu}{\Gamma},\frac{b}{\Gamma})$ of Eq.~(\ref{CurlH}) reduces to $\frac{(\mu/\Gamma)^2}{6\pi^2}$ which is independent of b and the result 
for 3D point node Dirac. It is clear that there are three regimes for $\mathcal{H}(\frac{\mu}{\Gamma},\frac{b}{\Gamma})$. A linear regime for $\mu<b$, a transition regime 
for $\mu>b$ during which $\mathcal{H}(\frac{\mu}{\Gamma},\frac{b}{\Gamma})$ evolves from a linear to quadratic dependence on $\frac{\mu}{\Gamma}$ and a final regime 
where the quadratic law $\mu^2/\Gamma^2$ holds. The size of $\mu$ relative to $b$ is critical in determining which regime is relevant for a particular value of $\mu$. 
Note that in Fig.~\ref{fig:Fig8} the three curves with finite $b$ all fall in the regime $\mu<b$ and so only the linear region of 
$\mathcal{H}(\frac{\mu}{\Gamma},\frac{b}{\Gamma})$ is probed. In Fig.~\ref{fig:Fig9} however we have chosen $b$ such that the dotted blue curve (clean limit) ranges over 
all three regimes involved in $\mathcal{H}(\frac{\mu}{\Gamma},\frac{b}{\Gamma})$ as $\mu/\Gamma$ ranges from $0$ to $10$. Below $\mu/\Gamma=1$ we are in the linear 
regime (dotted red curve) and above $8$ in the quadratic regime (solid black curve) with a transition region from linear to quadratic between these two extremes. For the 
dashed magenta curve however we do not show $\mathcal{H}(\frac{\mu}{\Gamma},\frac{b}{\Gamma})$ but rather have chosen to show only the linear dependence even outside its 
range of validity. It is clear that significant deviations from linearity for large values of $\mu/\Gamma$ enter only for $\mu/\Gamma\ge 8$. Comparing the case 
of finite $\Gamma$ with its clean limit (dotted curves) we see that the largest effect of finite $\Gamma$ is in the region $\mu/\Gamma\le 2$ where carrier 
scattering changes the linear law to a quadratic law.

We now turn to the small $\mu$ limit of $\sigma^{DC}(\mu)$ and obtain analytically the $\mu^2$ law seen in Fig.~\ref{fig:Fig8}. In fact it is convenient to return to 
Eq.~(\ref{DCIBFinal}) and Eq.~(\ref{DCDFinal}) and include at the same time finite temperature effects. We take $\mu/\Gamma$ and $T/\Gamma<1$ in which case 
it is justified to expand the integrand of Eq.~(\ref{DCIBFinal}) and Eq.~(\ref{DCDFinal}) to second order in $\omega$ dropping all higher order terms. After straightforward 
algebra we get that the D.C. conductivity takes the form
\be
\sigma^{DC}(\mu,T)= \frac{e^2}{\hbar^2 v_{F}} \lp\frac{\Gamma^2}{\pi^3}\rp \lb A + B(\mu^2+\frac{\pi^2 T^2}{3}) \rb,
\ee
where
\be
\label{A}
A=\sum_{s'}\int^{\infty}_{0}\hspace{-0.4cm} \rho d\rho \int^{\infty}_{0}\hspace{-0.4cm} dk_{x} \lp\frac{1}{\Gamma^2 + k^2_{x} +\lp \rho + s'b \rp^2}\rp^2.
\ee
and $B=B^{IB}+B^{D}$
with
\bea
\label{BIB}
&& B^{IB} = \sum_{s'}\int^{\infty}_{0}\hspace{-0.4cm} \rho d\rho \int^{\infty}_{0}\hspace{-0.4cm} dk_{x} \lp 1-\frac{k^2_x}{k^2_{x}\hspace{-0.1cm}+\hspace{-0.1cm}\lp \rho\hspace{-0.05cm} + \hspace{-0.05cm}s'b \rp^2}\rp \times \nonumber \\
&& \hspace{-0.3cm}\lb \frac{1}{\lp\Gamma^2 \hspace{-0.1cm}+\hspace{-0.1cm} k^2_{x}\hspace{-0.1cm}+\hspace{-0.1cm}\lp \rho \hspace{-0.05cm}+\hspace{-0.05cm} s'b \rp^2\rp^3} -\frac{2\Gamma^2}{\lp\Gamma^2 \hspace{-0.1cm}+\hspace{-0.1cm} k^2_{x}\hspace{-0.1cm}+\hspace{-0.1cm}\lp \rho\hspace{-0.05cm}+\hspace{-0.05cm}s'b \rp^2\rp^4}\rb 
\eea
and 
\bea
\label{BD}
&& B^{D} = \sum_{s'}\int^{\infty}_{0} \hspace{-0.4cm}\rho d\rho \int^{\infty}_{0}\hspace{-0.4cm} dk_{x} \lp \frac{k^2_x}{k^2_{x}\hspace{-0.1cm}+\hspace{-0.1cm}\lp \rho\hspace{-0.05cm}+\hspace{-0.05cm}s'b\rp^2}\rp \times \nonumber \\
&& \hspace{-0.3cm}\biggl[\frac{5}{\lp\Gamma^2\hspace{-0.1cm}+\hspace{-0.1cm} k^2_{x}\hspace{-0.1cm}+\hspace{-0.1cm}\lp \rho \hspace{-0.05cm}+\hspace{-0.05cm}s'b\rp^2\rp^3}\hspace{-0.05cm}-\hspace{-0.05cm}\frac{6\Gamma^2}{\lp\Gamma^2 \hspace{-0.1cm}+\hspace{-0.1cm} k^2_{x}\hspace{-0.1cm}+\hspace{-0.1cm}\lp \rho\hspace{-0.05cm}+\hspace{-0.05cm} s'b \rp^2\rp^4} \biggr].
\eea

These integrals can all be done analytically to yield
\bea
\label{SigmaDC}
&& \sigma^{DC}(\mu,T)= \frac{e^2}{2\pi^2\hbar^2 v_{F}}\biggl [ \sqrt{\Gamma^2+b^2} + \frac{1}{3\Gamma^2}\biggl\{\sqrt{\Gamma^2+b^2}\nonumber \\
&& -\frac{\Gamma^2 b^2}{2(\Gamma^2+b^2)^{3/2}} \biggr\} \times \{\mu^2 + \frac{\pi^2 T^2}{3}\}\biggr].
\eea
The limit of $b=0$ corresponds to 3D point node Dirac. The D.C. conductivity reduces to 
\be
\label{SigmaDC-Clean}
\sigma^{DC}(\mu,T)= \frac{e^2 \Gamma}{2\pi^2\hbar^2 v_{F}} \lb 1 + \frac{1}{3} \{\frac{\mu^2}{\Gamma^2}+\frac{\pi^2}{3} \frac{T^2}{\Gamma^2}\}\rb.
\ee
which is a function of $\mu/\Gamma$ and $T/\Gamma$ valid for both these variables less than one because of our expansion in $\omega$ to second order only. 
The D.C. conductivity is further linearly proportional to $\Gamma$. Expression (\ref{SigmaDC-Clean}) agrees with the previous work \cite{Nicol} in this limit. For a 
general $b$ and $\Gamma$ we get
\be
\label{SigmaDC-Clean1}
\sigma^{DC}(\mu,T)= \frac{e^2 \sqrt{\Gamma^2+b^2}}{2\pi^2\hbar^2 v_{F}} \lb 1 + \frac{h}{3} \{\frac{\mu^2}{\Gamma^2} + \frac{\pi^2}{3} \frac{T^2}{\Gamma^2}\}\rb,
\ee
with 
\be
\label{h}
h=1- \frac{\Gamma^2 b^2}{2(\Gamma^2+b^2)^2},
\ee
which is always near 1 in value. Its minimum is at $b=\Gamma$ where it is $\frac{7}{8}$, reduced from one by $\sim 12\%$. While the $\mu=T=0$ value of $\sigma^{DC}$
is very different in the nodal loop case from the 3D Dirac point node case, the first going like $b$ while the second goes like $\Gamma$, it has the same 
$\mu/\Gamma$ and $T/\Gamma$ dependence. T coefficient of this dependence is however always close to that of the point node Dirac.

Other important transport coefficients can also be calculated for $\frac{\mu}{\Gamma},\frac{T}{\Gamma} <1$. The formula for the conductivity is,
\be
\label{SigmaN}
\sigma^{tot}=\frac{e^2}{\hbar^2 v_{F}} \frac{\Gamma^2}{\pi^3} \int^{+\infty}_{-\infty}\hspace{-0.6cm}d\omega \lp- \frac{\partial f}{\partial \omega}\rp \lb A + B \omega^2\rb.
\ee

To get thermal conductivity($\kappa$) we drop the $e^2$ and add a factor of $(\frac{\omega-\mu}{T})^2$ which gives $\frac{\kappa_{22}}{T}$. The total thermal 
conductivity has a thermopower correction and reads,
\be
\frac{\kappa}{T}=\frac{\kappa_{22}}{T} - \frac{e^2 \kappa^2_{12}}{\sigma^{DC}},
\ee
where the formula for the coefficient $\kappa_{12}$ is given as in Eq.~(\ref{SigmaN}) without the $e^2$ and a factor $(\frac{\omega-\mu}{T})$ added to the integral of 
the energy integration over $\omega$. In terms of $\kappa_{12}$ the thermopower $S$ is given by,
\be
\label{Thermopower}
S = \frac{e\kappa_{12}}{\sigma^{DC}},
\ee
Another quantity often discussed is the Lorentz number ($L$) of the Wiedemann-Franz law. By definition, 
\be
\label{L}
L= \frac{\kappa}{T\sigma^{DC}}.
\ee
It is straightforward to obtain explicit results for the above coefficients using algebra closely related to that of Appendix B of Ref.~[\onlinecite{Nicol}]. The 
results are,
\be
\frac{\kappa_{22}}{T}= \frac{\sqrt{\Gamma^2+b^2}}{2\pi^2\hbar^2 v_{F}}\lb 1 + h \{ \frac{1}{9}\frac{\mu^2}{\Gamma^2} + \frac{7\pi^2}{45} \frac{T^2}{\Gamma^2}\}\rb,
\ee
\be
S=\frac{2\pi^2}{9e} \frac{h(\mu/\Gamma)(T/\Gamma)}{1+h\{\frac{1}{3}\frac{\mu^2}{\Gamma^2} +\frac{\pi^2}{9} \frac{T^2}{\Gamma^2}\}},
\ee
\be
L= \frac{\pi^2}{3e^2} \lb \frac{1+h \{\frac{1}{3}\frac{\mu^2}{\Gamma^2} + \frac{7\pi^2}{15} \frac{T^2}{\Gamma^2}\}}{1+h \{\frac{1}{3}\frac{\mu^2}{\Gamma^2} + 
\frac{\pi^2}{9} \frac{T^2}{\Gamma^2}\}}\rb,
\ee
with a correction to $L$ for thermopower of the form $S^2$. As we have seen $h=1$ for the point node Dirac case. For the nodal loop case $h$ is somewhat reduced going 
like $\lp1-\frac{1}{2}\frac{\Gamma^2}{b^2}\rp$ for $\Gamma/b<1$. When the disorder scattering rate becomes large as compared with $b$ we find that $h$ is reduced 
by $12\%$ at $\Gamma=b$ after which it again rises towards one. These results are only valid for $\mu/\Gamma$ and $T/\Gamma$ $<1$ near charge neutrality.

The approach to the minimum D.C. conductivity for $T=\mu=0$ but with finite $\Omega$ as $\Omega\rightarrow 0$ is also of interest. It takes on the form of 
Eq.~(\ref{SigmaDC-Clean1}) with the second term in the square bracket replaced by $\frac{1}{36(1+\frac{b^2}{\Gamma^2})^2}\lb 4\frac{b^4}{\Gamma^4}+6\frac{b^2}{\Gamma^2}+7\rb$. 
The expansion of the functions  $\Y(\bmu,\bOmega,\bb,\bepsilon_{s'}) $ of Eq.~(\ref{Special-function-interband}) and $\mathfrak{H}(\bmu,\bOmega,\bb,\bepsilon_{s'}) $ of 
Eq.~(\ref{Special-function-intraband}) in powers of $\bOmega$ and retaining terms to $\bOmega^2$ only gives,
\bea
&& \frac{\sigma(T=0,\bOmega)}{\Gamma}= \frac{e^2}{\hbar^2 v_{F}} \frac{\sqrt{\Gamma^2+b^2}}{2\pi^2} \times \nonumber \\
&& \lb 1+ \frac{1}{36(1+\frac{b^2}{\Gamma^2})^2}\lp 4\frac{b^4}{\Gamma^4}+6\frac{b^2}{\Gamma^2}+7\rp \frac{\Omega^2}{\Gamma^2}\rb,
\eea
where we needed to evaluate the integral,
\be
\frac{e^2}{2\pi^3 \hbar^2v_{F}}\sum_{s'}\hspace{-0.2cm}\int^{\infty}_{0}\hspace{-0.4cm} d\bk_{x}\hspace{-0.2cm}\int^{\infty}_{0}\hspace{-0.4cm}\brho d\brho \frac{4\bepsilon^2_{s'}-2\bk^2_{x}-1}{\lp 1+\bepsilon^2_{s'}\rp^4}=
\frac{4\frac{b^4}{\Gamma^4}+6\frac{b^2}{\Gamma^2}+7}{72\pi^2(1+\frac{b^2}{\Gamma^2})^{\frac{3}{2}}}. 
\ee

\section{Conclusion}
\label{sec:VI}

We have considered the effect of a finite scattering rate ($\Gamma$) on the finite frequency ($\Omega$) electromagnetic response of a nodal loop semimetal and on its D.C.
electrical conductivity, thermal conductivity, thermopower(or Seebeck coefficient) and Lorentz number (or Wiedemann Franz law). For the dynamic optical conductivity at 
zero temperature $\sigma(T=0,\Omega)$ several regime arises as a function of $\Omega$ even when the chemical potential is set equal to zero. In the limit of 
$\Omega\rightarrow 0$ which corresponds to the minimum D.C. ($\sigma^{DC}$) conductivity we find $\sigma^{DC}=\frac{e^2 \sqrt{\Gamma^2+b^2}}{2\pi^2\hbar^2 v_{F}}$ where 
$b$ is the radius of the nodal ring in energy units, $e$ is the electron charge, $\hbar$ the Plank's constant and $v_{F}$ the Fermi velocity. This reduces to the known 
result for 3D point node Dirac when $b=0$ and to the clean limit result for a nodal semimetal when $\Gamma=0$. In this last instance the correction for finite $\Gamma$ 
is of the order of $\Gamma^2/b^2$. The approach to the minimum D.C. conductivity obeys a $\Omega^2/\Gamma^2$ law with coefficient 
$\frac{4\frac{b^4}{\Gamma^4}+6\frac{b^2}{\Gamma^2}+7}{36\pi^2(1+\frac{b^2}{\Gamma^2})^2}$. This result agrees with the point node Dirac case of 
Ref.~[\onlinecite{Nicol}] when $b=0$. At frequencies a few times $\Gamma$ but smaller than $2b$ we find a constant interband background as in graphene of height 
$\frac{e^2}{\hbar^2 v_{F}} \frac{b}{16}$ provided $b>>\Gamma$ so that the ratio of $\frac{\sigma^{DC}}{\sigma^{BG}}=\frac{8}{\pi^2}$ and is exactly the same as one would 
get for graphene in the same constant $\Gamma$ approximation as used here. 
Of course both $\sigma^{DC}$ and $\sigma^{BG}$ are themselves different in that they both involve material parameters namely $v_{F}$ and $b$ while in graphene 
these drop out entirely of both properties. In the limit of $b\rightarrow 0$ the constant background loses its integrity and the conductivity $\sigma(T=0, \Omega)$ 
evolves toward its behavior in 3D Dirac point node semimetal. For $\Omega>2 b$, $\sigma(T=0, \Omega)$ again takes on the characteristic linear in $\Omega$ dependence and 
in fact the parameter $b$ completely drops out.

For finite value of the chemical potential $\mu$ a Drude like response is obtained in the regime $\Omega \le 2\mu$ with width related to the scattering rate $\Gamma$. In 
the pure case ($\Gamma=0$) the region up to $\Omega=2\mu$ would have zero conductivity but now this region shows only a depressed conductivity because of the disorder 
scattering. Beyond $\Omega\simeq 2\mu$ the flat background of the $\mu=0$ case is recovered if $\mu< b$ while we recover the linear dependence of the 3D point 
node Dirac case if $\mu>b$ with some smearing in the transition region around $\Omega=2\mu$. Finite chemical potential also affects the behavior of D.C. properties. For 
the electrical conductivity we find that provided $\mu/\Gamma$ (or temperature $T/\Gamma$) is less than $1$ the approach to charge neutrality is quadratic in
$\mu/\Gamma$($T/\Gamma$). We have also considered the limit of large $\mu$ at $T=0$ and find the quadratic behavior characteristic of the dirty limit $\mu^2/\Gamma^2$ 
approach to charge neutrality and then gradually goes into a linear law $\mu/\Gamma$ characteristic of a 2D graphene like system when $b$ is much larger than $\mu$ and 
this changes to a $\mu^2/\Gamma^2$ law for $\mu>>b$.

Finally the D.C. thermal conductivity is found to vary as $\frac{\sqrt{\Gamma^2 + b^2}}{v_{F}}$ when $\mu=0$, $T=0$ and have a $\mu^2/\Gamma^2$ or $ T^2/\Gamma^2$ 
correction for $\mu/\Gamma$, $T/\Gamma$ finite but smaller than one. The coefficient of these quadratic dependence are only slightly modified from the $b=0$ point node 
case. The maximum correction for finite $b$ is of order $12\%$. This implies that the Wiedemann-Franz law is only very slightly change from its value for $b=0$ (point 
node case). A similar situation holds for the thermopower. We hope that our calculations will stimulate experimental studies of the optical and transport properties of 
nodal loop semimetals. In particular A.C. spectroscopic data can provide a wealth of valuable information on the dynamics of the charge carrier. Several such studies 
already exist for related systems of Dirac and Weyl semimetals \cite{Chen, Sushkov, Xu, Neubauer, Chinotti}.

\subsection*{Acknowledgments}
Work supported in part by the Natural Sciences and Engineering Research Council of Canada (NSERC) and by the Canadian Institute for Advanced Research (CIAR).

\appendix

\end{document}